\documentclass[fleqn,usenatbib]{mnras}

\usepackage{newtxtext,newtxmath}

\usepackage[T1]{fontenc}
\usepackage{ae,aecompl}

\usepackage{graphicx}	
\usepackage{amsmath}	
\usepackage{amssymb}	

\usepackage{gensymb}
\usepackage{booktabs}
\usepackage{subfigure}
\usepackage{adjustbox}
\usepackage{placeins}
\usepackage[dvipsnames]{xcolor}
\usepackage{hyperref}
\usepackage{cleveref}
\usepackage{listings}

\newcommand{\gaia}{\emph{Gaia}}%
\newcommand{\gaiadrtwo}{\gaia\ DR2}%
\newcommand{\palfive}{\ensuremath{\rm{Pal\;5}}}
\newcommand{\panstarrs}{\rm{Pan-STARRS}}
\newcommand{\nbody}{$N$-body simulation}
\newcommand{\nbodys}{$N$-body simulations}

\newcommand{\pmra}{\mu_{\alpha^*}}
\newcommand{\pmdec}{\mu_{\delta}}

\makeatletter
\newcommand{\@todonotes@enable}{1}
\newcommand{\@todonotes@inline}{1}
\makeatother
\input{notes.tex}

\title[\palfive\ in \gaiadrtwo]{An extended Pal 5 stream in \gaiadrtwo}

\author[Starkman, Bovy, \& Webb]{
Nathaniel Starkman\thanks{E-mail: n.starkman [at] mail.utoronto.ca},
Jo Bovy, 
\& Jeremy J.  Webb
\\
Department of Astronomy and Astrophysics, University of Toronto, 50 St.  George Street, Toronto, Ontario, M5S 3H4, Canada
}
\date{Accepted XXX.  Received YYY; in original form ZZZ}

\pubyear{2019}

\begin{document}
\label{firstpage}
\pagerange{\pageref{firstpage}--\pageref{lastpage}}
\maketitle

\begin{abstract}
    We present the results of a detailed search for members of the Pal 5 tidal tail system in Gaia Data Release 2 (DR2).  Tidal tails provide a sensitive method for measuring the current and past gravitational potential of their host galaxy as well as for testing predictions for the abundance of dark matter subhalos.  The Pal 5 globular cluster and its associated tails are an excellent candidate for such analysis; however, only ${\sim}23 \degree$ of arc are currently known, with in particular the leading tail much shorter than the trailing.  Using Gaia DR2 and its precise astrometry, we extend the known extent of the Pal 5 tail to ${\sim}30 \degree$, $7$ degrees of which are newly detected along the leading arm.  The detected leading and trailing arms are symmetric in length and remain near constant width.  This detection constrains proposed models in which the Galactic bar truncates Pal 5's leading arm.  Follow-up spectroscopic observations are necessary to verify the candidate stream stars are  consistent with the known tidal tails. If confirmed, this Pal 5 stream extension opens up new possibilities to constrain the Galactic potential.
\end{abstract}





\section{Introduction} 
\label{sec:introduction}

	The tidal field of a globular cluster's host galaxy plays an important role in its dissolution, with the outer regions being more strongly affected by tides than the inner cluster regions \citep{Renaud18}.  In the approximation of a smooth Galactic mass distribution, the physical boundary of a globular cluster (GC) can be taken to be the surface along which the external tidal field and the cluster's potential are equal \citep{Renaud11}.  The L1 and L2 Lagrange points mark the locations where this surface intersects with the cluster's radial position vector.  Through internal processes, like relaxation and binary interactions, and external processes, like tidal heating and shocking, the anisotropic local tidal field results in stars primarily escaping through these Lagrange points \citep{Ross:1997fv,Fukushige:2000dt}.  As the cluster dissolves, leading and trailing tidal stellar tails will form in this way and these are typically referred to as ``stellar streams''.

	One of the most useful stellar streams to study is that of the globular cluster Palomar 5 (\palfive; \citealt{Odenkirchen:2001}).  The \palfive\ cluster lies at a sky position of $(\alpha,\delta) \approx (229\degree, -0.11\degree)$ and at a distance of $\sim 23 \, \rm{kpc}$ \citep{Odenkirchen:2003}.  \palfive's tidal tails are both thin and long, extending more than 10 kpc in the most sensitive detection so far \citep{Ibata2017}.  Observations of the \palfive{} tidal tail over the years, starting with the first detection by \citet{Odenkirchen:2001}, who found $\sim 10 \degree$ total of tidal tail, has consistently found the trailing arm to be longer than the leading arm.  \citet{Grillmair:2006} increased the detection limits to about $16 \degree$ along the trailing arm and $6 \degree$ along the leading arm using SDSS data \citep{McCarthy2006}, but detection of the leading arm was limited by the edge of the SDSS imaging survey footprint.  Subsequent studies have reproduced or extended these findings with different datasets.  \citet{Carlberg:2012bi} claims an extension of the trailing arm to nearly $23 \degree$ using SDSS DR8 data.  Again, the detection was limited by the survey footprint on the leading edge.  \citet{ibata2016} used Canada French Hawaii Telescope (CFHT) data to recover $19 \degree$ of stream, mainly along the trailing arm.  Finally, \citet{Bernard:2016} used \panstarrs\ data to recover the $\sim 22 \degree$ of stream from \citet{Grillmair:2006}, including the truncation of the leading arm at $\sim 8 \degree$.  The \panstarrs\ data extends to lower declination than the SDSS data and for the first time, the truncation in the leading arm was considered to be an intrinsic feature of the stream rather than an observational effect.  In total, only ${\sim}23 \degree$ have been conclusively detected, despite various attempts.  This detection is asymmetric -- ${\sim8} \degree$ of leading arm, ${\sim15} \degree$ of trailing arm --  favoring the trailing arm.

	From simulations of the formation of the \palfive\ tidal stream, we expect approximately symmetric tails \citep{Dehnen:2004eza} and the leading arm truncation is therefore puzzling.  We furthermore expect the tails to be longer than the observed extent, because the combined mass function of Pal 5 and its tail has a slope of ${\sim}0.6$ \citep{Grillmair:2006}.  This value is far from the believed primordial initial mass function, suggesting more stars have been lost than can currently be accounted for.  The missing stars should therefore compose an extended and symmetric tidal tail \citep{Koch2004}.

	To explain the leading-arm truncation, \cite{Pearson:2017} propose that the Galactic bar can induce stream asymmetry for certain values of the bar's pattern speed.  They additionally predict the leading arm should reappear further South.  However, it remains possible that the asymmetry in the detected tails may be the result of observational limitations: the leading arm more strongly overlaps with the bulge, making its detection more difficult.

	A data-set with a footprint and depth that allows one to search beyond previous surveys presents an enormous opportunity to extend the \palfive\ tidal tails and constrain the interaction history of the \palfive\ cluster with the Galaxy.  The Gaia Data Release 2 data-set (\gaia\ DR2; \citealt{GaiaCollaboration:2018}) fulfills the above two requirements.  \gaia\ is an all-sky survey comprising over 1.3 billion stars out to distances exceeding 23.5 kpc (the approximate distance of \palfive).  Moreover, unlike previous surveys, \gaia{} has parallax and proper motion measurements for most of its stars, which are useful for identifying a spatially and kinematically-continuous tidal tail.  One drawback of the \gaia\ data is the photometric uncertainty, which is relatively large compared to other surveys, especially at the magnitudes of detectable \palfive\ stars.  However, by cross-matching \gaiadrtwo{} to the \panstarrs{} survey \citep{Chambers:2016}, we are able to build a database with kinematics from \gaiadrtwo\ and precise \panstarrs\  photometry that combines the best aspects of both surveys.

    The purpose of this study is to use the combined \gaiadrtwo{} - \panstarrs{} dataset to search for \palfive's tidal tails.  The outline of this paper is  as follows: \autoref{sec:the_data} provides details of the data-sets used and the observed properties of the \palfive{} cluster, which form the basis of a set of filters applied in \autoref{sec:extended_stream} to the data-set.  These filters allow for stars that are member of the \palfive\ tidal tail to be identified, extending its length far beyond previous detection limits.  We summarize the results and their implications in Sections \ref{sec:results} and \ref{sec:conclusions}.



\section{Observations} 
\label{sec:the_data}

	The \palfive{} cluster, while undergoing tidal dissolution, is still well-defined \citep{Dehnen:2004eza} and can be easily identified in Gaia DR2 as an overdensity of points in right ascension and declination ($\alpha$, $\delta$).  Proper motions and the cluster's locus in the colour-magnitude diagram (CMD) help constrain cluster membership even further.  As previously discussed, stars which escape \palfive{} are expected to populate tidal tails that precede and proceed the cluster along its orbit.  The following data cuts that reveal the cluster will likewise distinguish the tidal tail.

    \subsection{Gaia Data} 
    \label{sec:gaia_data}

			\begin{figure}
				\centering
				\includegraphics[width=.8\columnwidth]{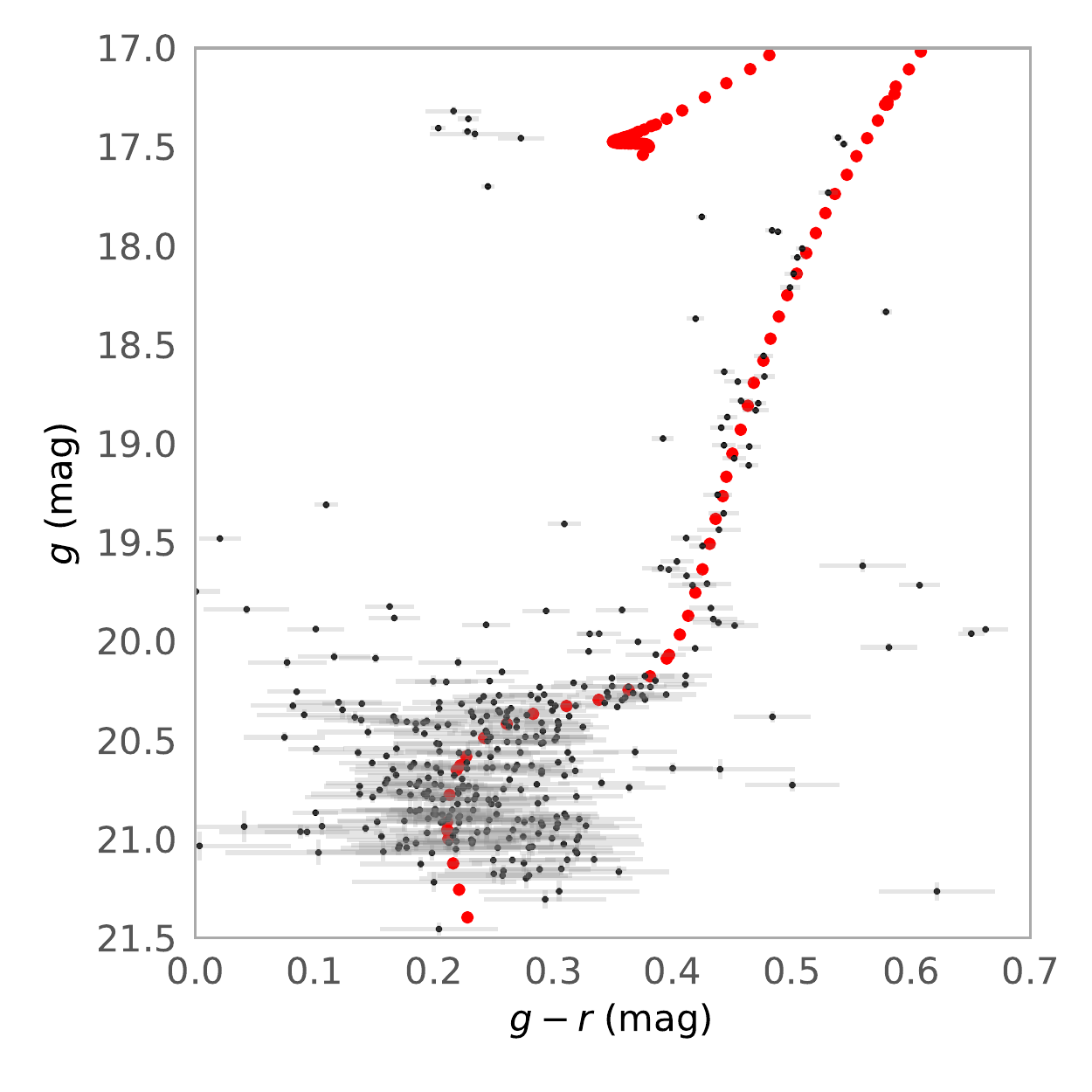}
			    \caption{CMD of all stars with $\varpi < 1 \, \rm{mas}$ within 3 arc-minutes of the center of \palfive{}.  The locus of \palfive{} is strongly apparent with a well-populated giant branch and MSTO.  Though sparse, there is a distinguishable horizontal giant branch and collection of blue stragglers.  The latter are not used when constructing the CMD filter.}
			    \label{fig:pal5_progenitor_cmd}
			\end{figure}

     We obtained observational data from a \gaiadrtwo{}-\panstarrs\ cross-matched catalogue, available via the \gaia{} data portal \citep{GaiaCollaboration:2016, Luri:2018eu}.
        The code to perform this cross-match and subsequent analysis will be made publicly available \footnote{ \href{https://github.com/nstarman/Pal-5-in-Gaia-DR2}{https://github.com/nstarman/Pal-5-in-Gaia-DR2}}.
        To select stars in the cross-matched catalogue that could be members of the \palfive{} tails, we construct a sky-rotated coordinate system which locally linearizes the \palfive{} tail about a point.  The proper sky rotation for \palfive{} cannot be known \emph{a priori}; hence the sky-rotation matrix is defined by proxy using an N-body simulation of \palfive{} and its tidal tails.  We define the \palfive{} cluster-centered system with the coordinates $\phi_1$- - along the direction of the tails---and $\phi_2$---perpendicular to the tail.  We select the rotation such that, moving from the cluster, the tails curve up toward positive $\phi_2$.  The details of the coordinate system are provided in Appendix \ref{sec:constructing_the_coordinate_system}.
        

	\subsection{Parallax Cut} 
	\label{sub:parallax_cut}
	
	    The \palfive{} cluster is thought to be approximately 23 kpc away \citep{Odenkirchen:2003,Bovy:2016cx}.  In \gaiadrtwo{}, main-sequence turn-off (MSTO) stars can be seen to this distance near \gaia's faint limit, but the data do not have precise parallax measurements in this regime \citep{GaiaCollaboration:2018}.  Consequently, we only constrain parallaxes $\varpi$ to be less than $1\,\mathrm{mas}$ to cut all data closer than 1 kpc.  More stringent cuts are too restrictive given the errors.  With this cut the \palfive\ GC is clearly present, but the tail is indistinguishable from foreground or background objects.  Further selection procedures are needed.  Therefore we next examine the \palfive\ GC's proper motion and locus in the colour-magnitude diagram to determine the filters necessary to extract the extended tails.


	\subsection{Proper Motion Cut} 
	\label{sub:proper_motion}

		\begin{figure}
			\centering
			\includegraphics[width=.8\columnwidth]{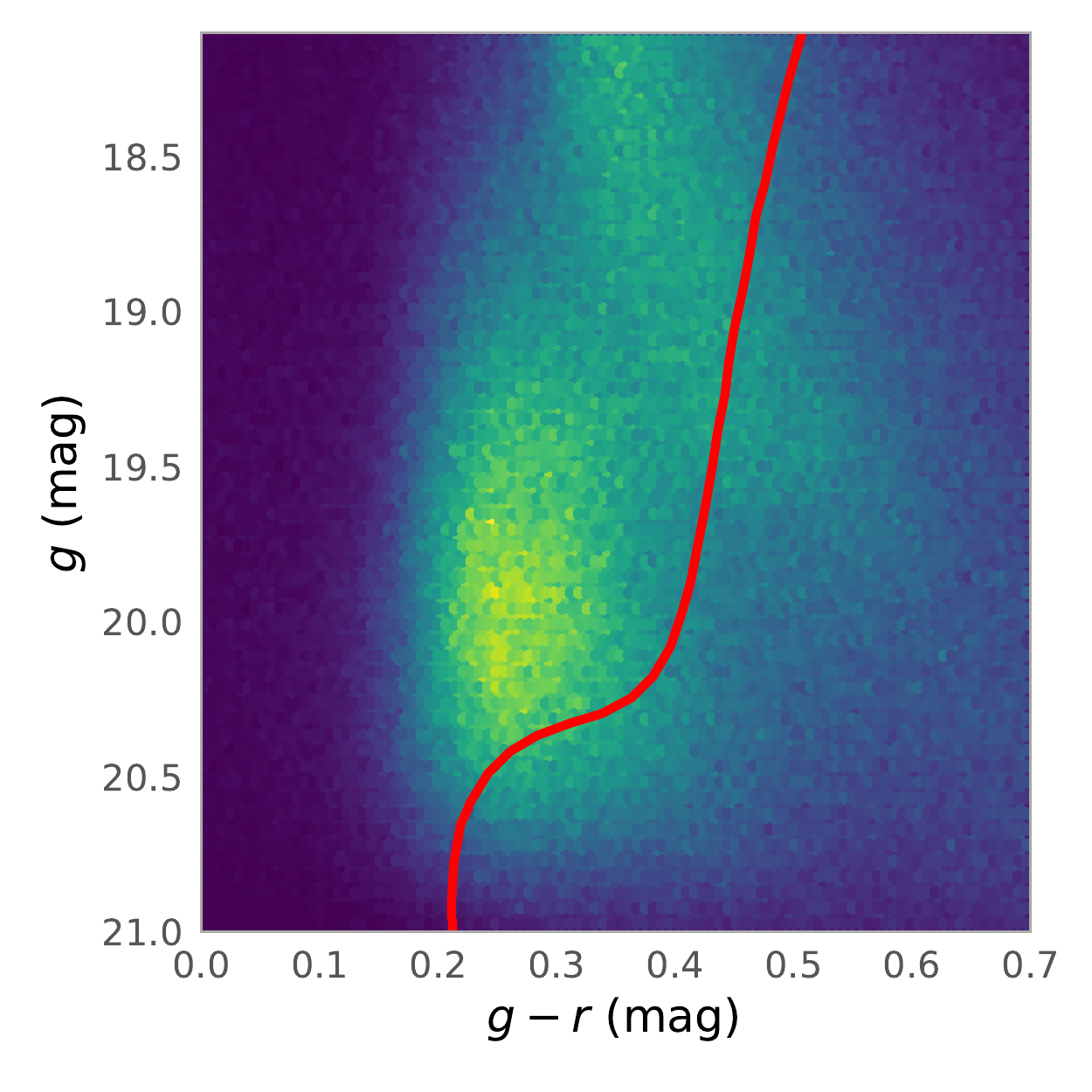}
		    \caption{The background distribution of stars in the \gaiadrtwo{} data window around the \palfive{} cluster.  The bulge is clearly visible in the histogram, coloured by the logarithm of the density.  Overlayed on the bulge is the best-fit isochrone.  The densest region of the bulge almost directly overlays \palfive's main sequence turn-off.
		    }
		    \label{fig:background_distribution}
		\end{figure}

		An advantage of \gaia{} over previous instruments is its ability to accurately measure stellar kinematics.  Pre-\gaia\ stream detections have generally been limited to photometry.  To remain coherent on the sky, stars in a stellar stream need to be close to co-moving and the kinematics of the tails are similar to that of the progenitor, with differences owing primarily to the tidal dissolution process.  As the process can be be modeled \citep[e.g.,][]{Kruijssen2009,Bovy:2014,Webb:2018dw,Balbinot2018}, examining the cluster's kinematics allows for tail stars to be identified amongst background stars not associated with \palfive{}.

		The kinematics of the cluster have been measured often, most recently by \citet{Vasiliev:2018it} using the latest \gaiadrtwo{} data.  Our findings agree with \citet{Vasiliev:2018it}: both in right ascension and declination, the proper motion distribution of stars within 2 projected tidal radii of \palfive{} is sharply peaked around $-2.7 \; (\rm{mas} \, \rm{yr}^{-1})$.  Although this peak suggests individual cluster members may be identified by only their proper motion and spatial clustering, at the distance of \palfive, the errors in \gaiadrtwo{} proper motion data preclude this direct approach.

		As will be explained in \hyperref[sub:stream_proper_motion]{Section \ref*{sub:stream_proper_motion}} below, in order to isolate the tidal tail, we make a less restrictive proper motion cut by selecting a circle in right ascension and declination proper motion space that encompasses the \palfive\ tails' proper motion distribution and accounts for the significant errors in the proper motions.  The proper motion cut allows for the tails' location in the CMD to be identified.


	\subsection{Colour-Magnitude Cut} 
	\label{sub:prog_cmd}

        Complimentary to the proper-motion strategy to isolate the \palfive{} cluster and tail is a more traditional photometric selection, accomplished using the CMD of the cluster.  In \autoref{fig:pal5_progenitor_cmd} the cluster's locus is well defined in the CMD in the $g$ and $r$ \panstarrs{} bands.  These bands are optimal for detecting main-sequence stars (the majority of tail stars) and distinguishing tail-stars from foreground and background objects \citep{Ibata2017}.  A range of isochrones can be fit to \palfive{}'s CMD due to the age-metallicity-distance degeneracy in isochrone tracks.  As we are not intending to precisely distance or age \palfive{}, any isochrone fitting the data is sufficient.  We find that a PARSEC isochrone \citep{Bressan2012} with an age of $11.2$ Gyr and metallicity $\left[{\rm{Fe}}/{H}\right] = -3.072$ fits the locus of the \palfive\ stars well\footnote{The isochrone was generated using CMD v3.1 (\href{http://stev.oapd.inaf.it/cgi-bin/cmd_3.1}{http://stev.oapd.inaf.it/cgi\-bin/cmd\_3.1}), with the \panstarrs{} photometry and a \citet{Marigo:2008} isochrone track.}.  Placed at a distance of ${\sim}22 $ kpc, the isochrone tracks the MSTO and the giant branch.

		\begin{figure}
			\includegraphics[width=1\linewidth]{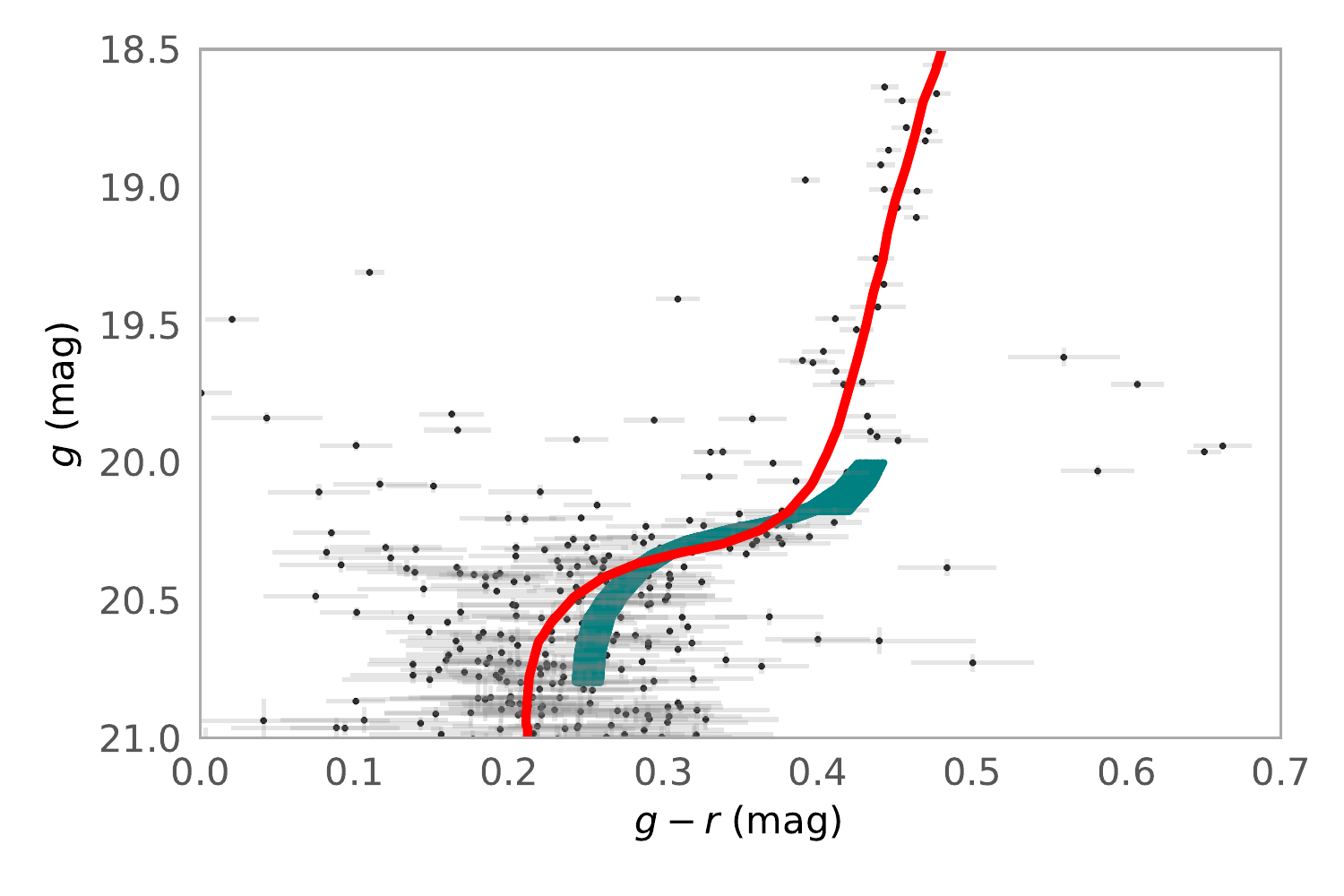}
		    \caption{The colour-magnitude cut used to find the extended \palfive{} tidal tails.  The black dots are stars with $\varpi < 1 \, \rm{mas}$ within 3 arc-minutes of the center of \palfive{}.  An isochrone is fit through the cluster.  The CMD cut (teal) is constructed based off this isochrone, and is limited to $20 < g  < 20.7$ mag.}
		    \label{fig:shifted_cmd}
		\end{figure}

		Past work on detecting the \palfive{} tidal tails has always fit a CMD-cut directly to the \palfive{} cluster (see, e.g., \citealt{ibata2016}).  This works well when using data-sets with deep photometry.  However, as is apparent in \autoref{fig:pal5_progenitor_cmd}, the main sequence (MS) of \palfive{} is poorly constrained in \gaiadrtwo{}, even with \panstarrs{}-crossmatched photometry.  This is because it is largely below \gaia's magnitude limit and fainter than a magnitude of ${\sim}21$, the $g$-band photometry is noise-dominated.  The noise is problematic as the majority of stars in both the cluster and tail are MS stars.  Excluding these stars leaves little data.  As a compounding problem, in \autoref{sec:extended_stream}, we demonstrate that the foreground, primarily from the Galactic bulge, dominates the CMD near \palfive\ in the region $18 \lesssim g \lesssim 20.7$ mag, cutting into the expected location of the tails' MSTO.  The significant amount of foreground stars precludes the standard application of an isochrone-based cut fit directly to the cluster.  A solution to this problem is presented in \hyperref[sub:stream_colour_magnitude_relation]{Section \ref*{sub:stream_colour_magnitude_relation}}.




\section{Identifying the Extended Stream} 
\label{sec:extended_stream}

	In this section, we describe how we adapt and apply the techniques used in \autoref{sec:the_data} to examine the \palfive\ cluster, to reveal a much-extended length of the \palfive\ tidal tails.

	\subsection{Parallax and Magnitude Cut} 
	\label{sub:distance_cut}

\begin{figure*}
	\centering
	\includegraphics[width=1.\textwidth]{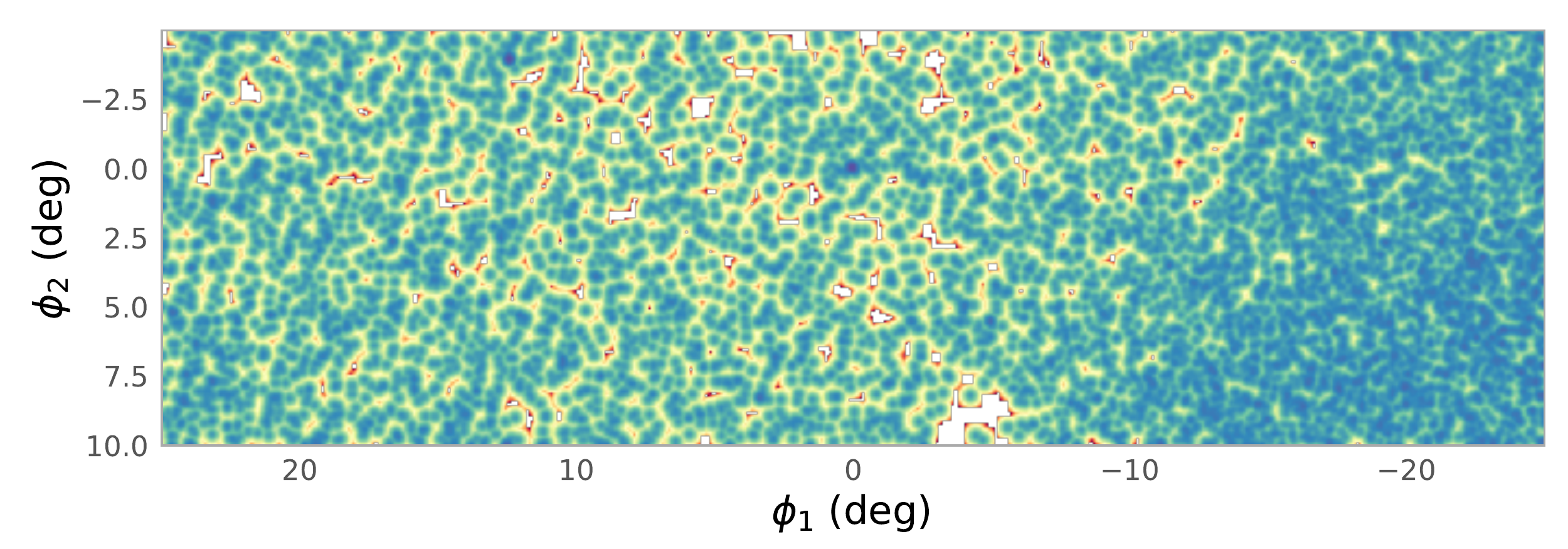}
    \caption{The \palfive\ data-set with all cuts--- parallax, proper motion, colour-magnitude, Gaussian filter---applied. The leading and trailing arms clearly extend from the progenitor at $(\phi_1,\phi_2) = (0,0)$ for $\sim15^\circ$ on both sides. The leading arm becomes indistinguishable past $\sim (15, 2.5)\degree$, while the trailing arm does so at $\sim (-12, 2.5)\degree$.}
    \label{fig:square_closeup_all_cuts}
\end{figure*}

		Like in \autoref{sec:the_data}, we apply a $\varpi < 1\,\mathrm{mas}$ cut to remove nearby stars from the dataset.  The photometric depth of the \gaiadrtwo{} - \panstarrs{} cross-match catalogue is approximately $g \sim 21$ for stars with $-0.2 \lesssim g-r \lesssim 0.6$, the range relevant for \palfive{} members.  As shown in \autoref{fig:pal5_progenitor_cmd}, at the distance of \palfive\ the majority of the main sequence lies below $g = 21$.  At $g \gtrsim 20.5$  the data becomes progressively more noise dominated.  Because few stars occupy the giant branch of the \palfive{} isochrone, we adopt the $g$-band filter $20 < g < 20.7$.  This range encompasses the MSTO and a limited portion of the main sequence, avoiding the noise-dominated measurements nearer $g {\sim}21$ mag.


	\subsection{Proper Motion Cut} 
	\label{sub:stream_proper_motion}

		While the kinematics of the \palfive{} cluster are fairly well constrained \citep{Vasiliev:2018it} and the tidal tail kinematics are similar to that of the cluster, predicting the tails' exact proper motion at distances far from \palfive{} is not trivial.  The tidal dissolution process may be modeled \citep[e.g.,][]{Webb:2014,Bovy:2014}, but it is an area of active research and different models predict different proper motion distributions.  Additionally, the choice of Galactic potential model has a large effect on the resulting stream kinematics \citep{Bovy:2016cx}.  We use an \nbody{}, described in \hyperref[sub:the_nbody]{Appendix \ref*{sub:the_nbody}}, to constrain the range of proper motions that tail stars could have.

		The proper motion cut is constructed by taking a circle in $\pmra$ and $\pmdec$ near the centroid of the $N$-body simulation's proper motion distribution.  The circle center is located at $(\mu_{\alpha^*}, \mu_{\beta}) = (-2.5, -2.5) \ \rm{mas}\,\rm{yr}^{-1}$, which better reflects the whole system as opposed to just the GC.  The selection radius is $3 \ \rm{mas}\,\rm{yr}^{-1}$ encompasses the data points at the extrema of the tails to around the 1-$\sigma$ level after accounting for the observational errors.


	\subsection{Colour-Magnitude Cut} 
	\label{sub:stream_colour_magnitude_relation}
	
\begin{figure*}%
	\centering
	\subfigure[Sky projection onto an orbit fit to the trailing arm][{Sky projection onto an orbit fit to the trailing arm}]{%
	\label{fig:trail_arm_projection}%
	\includegraphics[width=\columnwidth]{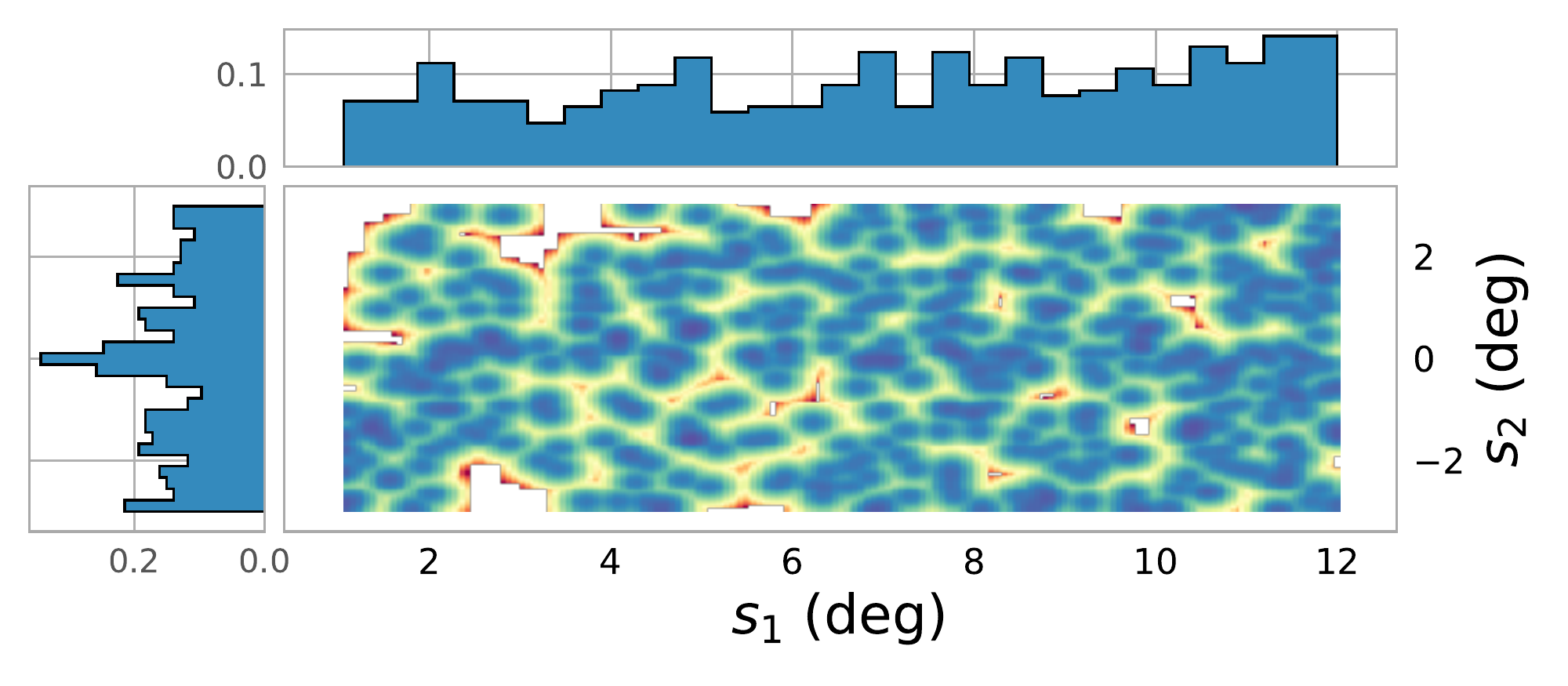}\textbf{}}%
	~
	\subfigure[Sky projection onto an orbit fit to the leading arm][{Sky projection onto an orbit fit to the leading arm}]{%
	\label{fig:lead_arm_projection}%
	\includegraphics[width=\columnwidth]{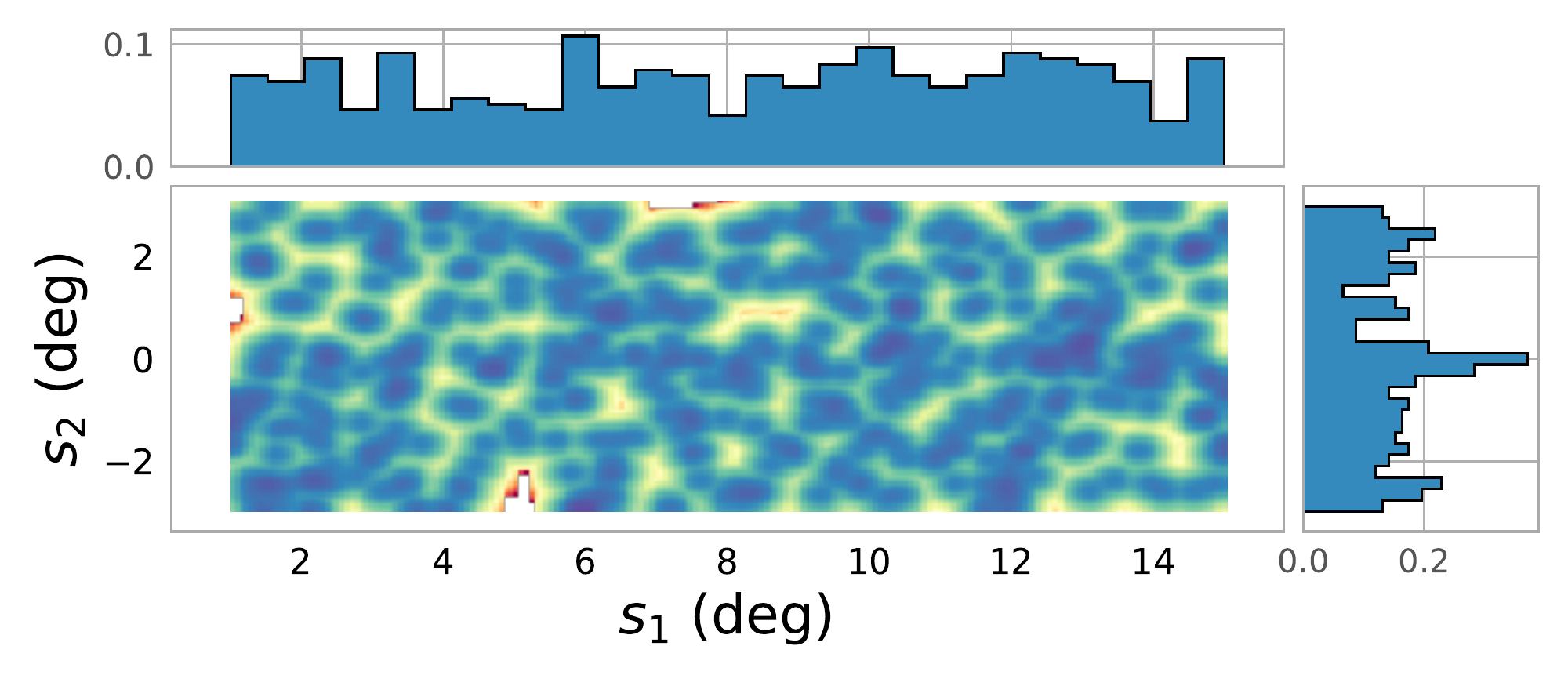}}%
	\\

	\caption[tidal tail projections]{These figures show the nearest $3\degree$ of the \palfive\ data-set projected along orbits fit to the tails.  The coordinate $s_1$ tracks the tail arc-length, and $s_2$ is the orthogonal distance from the tails, which are centered on $s_2 = 0$.  The left plot displays the leading arm of the \palfive{} tidal tail, starting a distance of $1 \degree$ from the cluster so as not to include any cluster stars.  The tail is visible for ${\sim}15 \degree$.  The right plot is that of the trailing arm.  On this side the tail is visible for at least $12 \degree$.}%
	\label{fig:projections}%
\end{figure*}

		Standard practice to isolate the \palfive{} cluster and its tidal tails is to select stars within some region surrounding \palfive's locus in the CMD using an isochrone \citep{Grillmair:2006,ibata2016}.  In \autoref{sub:prog_cmd}, we fit an isochrone to the cluster's CMD locus that could serve as the basis of such a filter.  However, the standard isochrone approach fails due to the dominance of the Galactic bulge, whose stars overlap with and far outnumber the stars in \palfive{}'s MSTO.  For deep photometric surveys, such as used in the two aforementioned papers, the CMD cut extends into \palfive{}'s main sequence where there is little contamination from the bulge.  However, at the brighter magnitudes accessible in \gaia, a significant population of Galactic bulge stars exists in a similar part of the CMD as \palfive's MSTO region.  A standard CMD cut around the best-fit isochrone therefore does not reveal the extended \palfive{} tails as it picks up too many bulge foreground stars.  The dominance of the bulge may be seen in \autoref{fig:background_distribution}.  Overlayed on the plot is the best-fit isochrone (\hyperref[sub:the_nbody]{Section \ref*{sub:prog_cmd}}).  The bulge distribution is most dense around $(g-r, g) {\sim}(.25, 20.2)$, abutting the top of the cluster's MSTO, where nearly all \gaia{}-observable stars live.

		Without an extended MS, the presence of the bulge precludes the standard isochrone approach.  The solution is to construct a CMD cut as far away as possible from the Galactic bulge locus, while still maintaining enough \palfive{} stars to observe the tails.  \autoref{fig:shifted_cmd} demonstrates this careful balancing act.  The red line is the isochrone described in \autoref{sub:prog_cmd} (with an age of $11.2 $ Gyr and a metallicity of $\left[{\rm{Fe}}/{H}\right] = -3.072$).  This is the isochrone about which a standard CMD cut would be constructed.  The width of the CMD cut would be set by the stellar overdensity and be thinnest at the MSTO.  Instead, the CMD shown in teal is used. This `shifted'-CMD is the region between two translations of the best-fit isochrone in colour-magnitude space, with an additional broadened region on the giant branch of the \palfive\ isochrone. The translation distance of the `shifted'-CMD from the best-fit isochrone is selected such that the bulge locus is avoided, while the majority of the MSTO remains within the selection. A CMD shifted further than this contains insufficient cluster and tails stars, while much closer CMDs fall prey to the Galactic bulge. The upper region of the `shifted'-CMD is manually broadened to track the stars with $\varpi < 1 \, \rm{mas}$ within 3 arc-minutes of the center of \palfive{}. These stars, shown in \autoref{fig:shifted_cmd}, do not appear to lie along the best-fit isochrone, necessitate the broadened selection in this region.

		While the \palfive{} tail has a distance gradient, the background does not.  Also, at ${\sim}23$ kpc, the colour-magnitude locus of the Galactic bulge lies at lower magnitude than the cluster isochrone.  Since the \palfive{} cluster is near apocenter at the other side of the Galaxy relative to us, the leading and trailing tails lie closer than the cluster and thus move progressively into the Galactic bulge's locus in the CMD the further along the tail one travels.  Nearer regions of the tidal tails have even stronger overlap than the cluster to the Galactic bulge.  However, as the tail shifts to lower $g$-magnitude it brings stars in the tails' MS above \gaia{}'s magnitude limit.  Therefore, disregarding the distance gradient and keeping the CMD cut static allows the true CMD along the tail to sweep through the CMD cut as defined around the cluster, and minimize the number of Galactic bulge stars selected by this cut.


	\subsection{Smoothing Kernel} 
	\label{sub:smoothing_kernel}

		To maximize the signal-to-noise of the tails star-count overdensities, we use a Gaussian smoothing kernel of ${\sim}0.1^\circ$, approximately the same angular width as the tails.  We motivate the smoothing width by examining the previously-detected tail lengths, such as those in \citet{ibata2016} and \citet{Grillmair:2006}.
		This single-kernel approach has two drawbacks.  First, the characteristic width of the \palfive{} tails is not precisely determined.  Both internal and external kinematics of the cluster are important to the tail width, as well as the precise potential of the Galaxy and the interaction history of \palfive{} with its environment.  Second, \nbodys{} of \palfive{}, such as from \autoref{sub:the_nbody}, predict a spatial broadening along the tail.  Therefore, there is no single characteristic tail spatial width.  However, as predicted by the N-body simulation, the tail remains thin for tens of degrees on either side of the cluster.  Also, because we aim to measure the tails' length and not their detailed properties, small deviations from a single width are not important.  For these reasons, despite theoretical drawbacks, the single-kernel approximation remains valid along the tails' detected length.  

	
	\subsection{Dust Foregrounds and Observational Cadence} 
	\label{sub:foregrounds_and_observational_cadence}

		We examine the data-set to ensure dust foregrounds and \gaia's observational cadence are not mistaken for the tidal tail.  Photometric dust extinctions in the \panstarrs{} $g$ and $r$ are obtained using the \verb!mwdust! package \citep{Bovy:2015br}.  Examining the dust maps reveals no structure in the dust to which the detected tail might be attributed.  Likewise, a hexagonal tiling pattern from \panstarrs{}, observable in some proper motion cuts, is not consistent with the tidal tail.  We also note that vertical striping from \emph{Gaia}'s observational cadence is nearly orthogonal to the tail and cannot explain the observation.  Finally, no systematic discussed in \citet{GaiaCollaboration:2018} aligns with the predicted (or past measured) tail path within at least 20 degrees of the cluster.
		Therefore, the observed tidal tail is not a dust or observation-strategy artifact--indicating it is a true measurement.



\section{Results} 
\label{sec:results}

    Applying the filters and processing from \hyperref[sec:extended_stream]{Section \ref*{sec:extended_stream}} reveals \palfive's tidal tails in \autoref{fig:square_closeup_all_cuts}.  On the trailing arm---in the negative $\phi_1$ direction---the arm becomes indistinguishable from background after ${\sim}12 \degree$.  On the opposite side, the leading arm extends ${\sim}15 \degree$, though there are hints it goes further (see \autoref{fig:square_closeup_all_cuts}), before likewise becoming indistinct.  Overall, this represents a ${\sim}27 \degree$ detection of the \palfive{} tidal tail and an extension of the leading arm by ${\sim}7 \degree$.

    Prior detections of \palfive{}'s tidal tail have mostly been along the trailing arm, starting from \citet{Odenkirchen:2003} to \citet{Bernard:2016}, as described in \autoref{sec:introduction}.  In total, only ${\sim}8 \degree$ has been detected along the leading arm, while we recover $15 \degree$.  Along the trailing arm, we find the stream along $12 \degree$, only slightly shorter than prior detections, which detect the trailing arm out to ${\sim}15 \degree$ \citep{Bernard:2016}.  The total length of the \palfive\ stream is therefore ${\sim}30 \degree$, symmetric around the progenitor.
    
    While many of the prior studies have deeper photometry than is available in \gaiadrtwo{}, none have as extensive kinematic nor parallax data.   These data have proved a crucial development to stream searches.  Stream-search algorithms, such as developed by \citet{Malhan:2018dp} and \citet{Palau2019}, have discovered many new streams too faint to detect using purely photometric techniques.  It is in this same vein that we present extensions to the \palfive{} tidal tail.  \gaiadrtwo{} photometry is not deep, but the extensive astrometric data permits more restrictive phase-space reductions than in prior data sets, and when combined with photometric and parallax restrictions, reveals an extended \palfive\ tidal tail.

	All the phase-space restrictions described in \hyperref[sec:extended_stream]{Section \ref*{sec:extended_stream}} are necessary to identify the \palfive{} tidal tails.  The 1 kpc parallax filter removes the majority of foreground objects.  The proper motion filter removes $81 \%$ of the remaining stars, most importantly objects such as the large M5 globular cluster, which lies within 2 degrees of the \palfive\ cluster.  The single most restrictive cut is that of the shifted-CMD, motivated in \hyperref[sub:stream_colour_magnitude_relation]{Section \ref*{sub:stream_colour_magnitude_relation}}.  This CMD cut removes 98 \% of the remaining stars in the selected \gaia{} window.  While many potential tail stars are cut, the CMD cut crucially removes most foreground objects, particularly the bulge stars which enshroud the tidal tail.  The CMD cut is particularly efficacious along the \palfive{} leading arm and permits a $15 \degree$ tail-length detection.
	
	Corroborating the visual detection of the tail are the projections of the \gaiadrtwo{} data on orbits hand-fit to align with the observed tidal tails.  These projections are shown in \autoref{fig:projections} for both the leading and trailing arms.  The orbits are only fit to the tail in $\alpha$ and $\delta$ and are not predictions of the actual orbital dynamics of the tidal tail; we only use them to provide a convenient parametrization of the path of the tails.  The positions and velocities of these hand-fit orbits are given in \autoref{tab:hand_fit_orbits}.  Stars within $3 \degree$ of the orbits are selected and represented in the on-stream coordinate system $(s_1, s_2)$---the arc length along and orthogonal distance to the stream, respectively.  The fits along $s_2$ confirm an enhancement at $s_2=0$, the location of the tails, above the background level of the star count near the tails.
	
	\begin{table}
    \caption{Positions and velocities for the orbits hand-fit to the sky positions of the tidal tails for the purpose of straightening the tails.  Projections of stars within 3 $\degree$ of these orbits are shown in \autoref{fig:projections}.}
    \label{tab:hand_fit_orbits}
    \resizebox{\linewidth}{!}{%
        \begin{tabular}{lllll}
            arm & $\alpha, \delta (\degree)$ & $d (\rm{kpc})$ & $\mu_{\alpha^*}, \mu_{\delta} (\rm{mas}\,\rm{yr}^{-1})$ & $v_r (\rm{km}\,\rm{s}^{-1})$ \\ \hline
            Leading & 229.022, -0.223 & 22.5 & -2.16, -2.23 & -40 \\
            Trailing & 229.022, -0.223 & 22.5 & -2.00, -2.18 & -50 \\
            \hline
        \end{tabular}%
    }
    \end{table}

\begin{figure}
	\centering
	\includegraphics[width=1.\columnwidth]{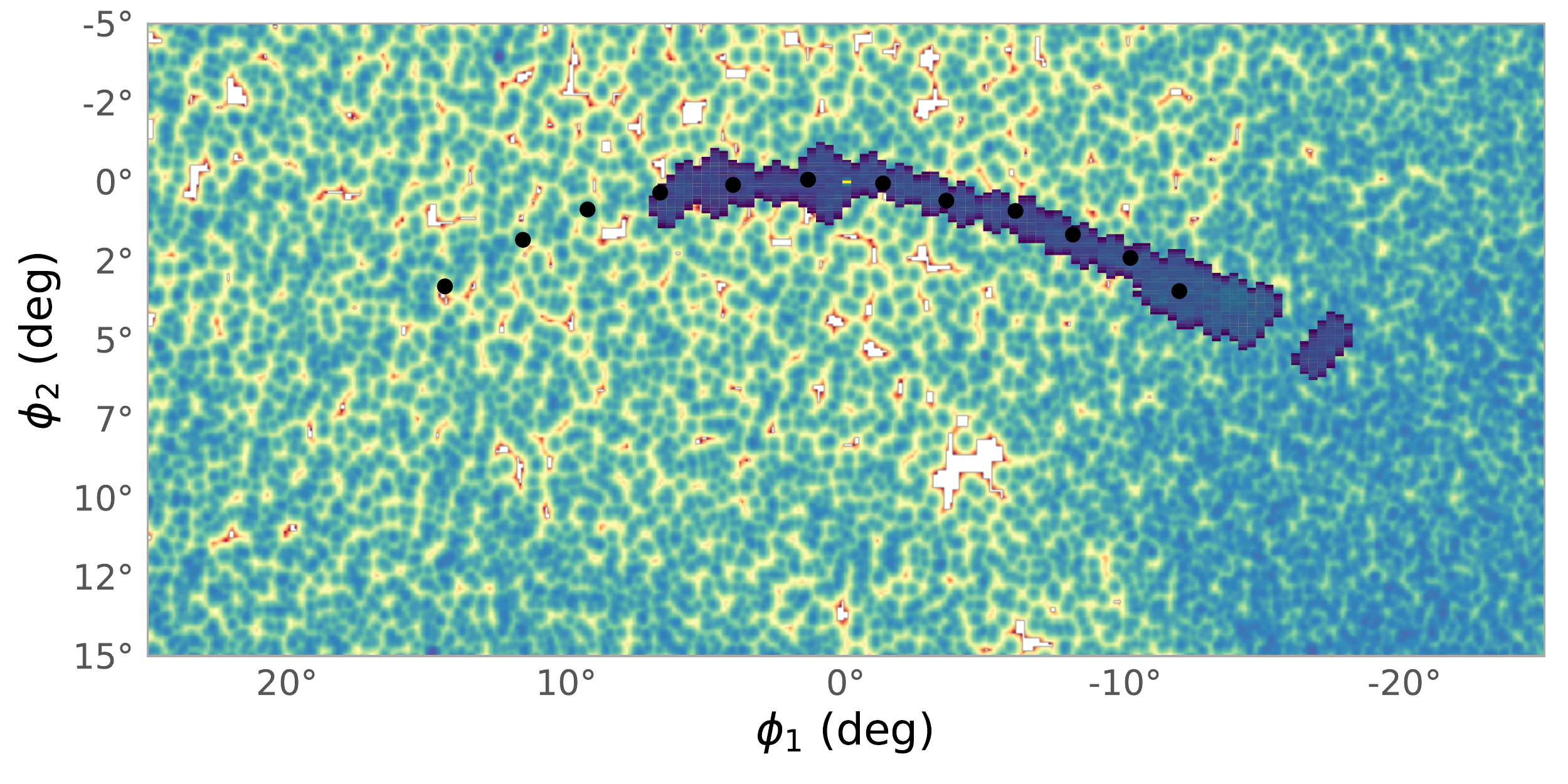}
    \caption{Comparison between our \palfive\ stream detection and previous work. We overplot the \citet{ibata2016} data on the \gaiadrtwo{} data-set of this paper.  Also shown are a number of black points which are placed along our observed tails to assist in comparing the two data-sets.  Over the region where the data-sets overlap, the tidal tails agree.}
    \label{fig:consistency_with_ibata}
\end{figure}

	We compute the significance of our new detection using Poisson anomaly detection on the binned star counts.  As the stream is spread over multiple bins, we calculate the likelihood of observing a set $S$ stream-bins, from $n$ bins total, with star counts $f$ and median $\bar{x}$ as 
	$$p = (n-2) \prod_{i\in{S}}{\left(1 - \rm{CDF}_{Poisson}(f_i, \bar{x})\right).}$$
	$S$ is the index set for bins in the stream, with number of elements $\geq s$.  The $(n-2)$ factor accounts for the fact that the anomaly could have been detected in any of the bins, except for the edge bins.  \autoref{fig:projections} shows a $3$ degree region surrounding each stream.  The leading arm and trailing arms have at least 2 stream bins, corresponding to a width of $\sim0.5\degree$ and consistent with previous studies \citep[e.g.,][]{Grillmair:2006}.  Using the data cuts previously described, from these bins, we achieve a {$> 3\, \sigma$} detection along each the trailing and leading arms. The significance of the newly-detected part of the leading arm, excluding the previously detected $8\degree$, is also $\gtrsim 3\,\sigma$.

	An important validation of our approach is that of consistency with prior work.  The dataset used in \citet{ibata2016} is publicly available on the \href{http://vizier.u-strasbg.fr/viz-bin/VizieR-2}{VizieR database}.  \autoref{fig:consistency_with_ibata} shows that the observed tidal tail exactly tracks with the results of \citet{ibata2016}.  We also checked against the results of \cite{Carlberg:2012bi} and find consistent results. 

\section{Conclusions} 
\label{sec:conclusions}

	To date, ${\sim}23 \degree$ of \palfive{}'s tidal tails have been detected, with the leading arm significantly shorter than the trailing arm \citep{Bernard:2016}.  As explained in \Cref{sec:introduction}, we expect longer and more symmetric tidal tails unless the \palfive{} stream has been significantly perturbed.

	There is a convergence of factors which make the \palfive{} tidal tail difficult to detect in \gaiadrtwo.  First, it is distant, requiring deeper photometry than is available in \gaiadrtwo{}.  Second, at this distance, the proper motions in \gaiadrtwo{} have fractional errors of order unity.  Third, the Galactic bulge overlays part of \palfive{}'s MSTO, the highest number-density area available in the CMD.  However, using a parallax, proper motion, and CMD cut, and applying Gaussian spatial smoothing, we report a detection of ${\sim}27 \degree$ of \palfive{}'s tidal tails.  Of this, ${\sim}7\degree$ are newly-detected extensions to the \palfive{} tails.  Combined with previous detections of the trailing tail, we find a total length of $\sim 30\degree$, $15\degree$ on each side of the Pal 5 cluster. Given the distance to \palfive{}, stars populating the region of the tail ${\sim}15\degree$ away escaped the cluster several Gyr ago \citep{Dehnen:2004eza}.  Given the length and width of the tail, the cluster has been continuously losing mass for several Gyr without a major disruption event.

	Most importantly, the extension we find lies entirely along the leading arm, which has historically been found to be significantly shorter than the trailing arm \citep{Odenkirchen:2001}.  Taking our newly-detected extension into account, the two arms are roughly equal in length.  This is in contrast to previous studies, many of which speculate that the leading arm is truncated compared with the trailing arm.  \citet{Pearson:2017} hypothesize that interactions with the bar curtail (or prevent) stars escaping into the leading arm and thus cause an underdensity (or gap) in the leading arm, starting at ${\sim}8 \degree$.  Our finding of symmetric tidal tails constrains this scenario.  As we are sample limited, we cannot rule out scenarios which reproduce our threshold density past ${\sim}8 \degree$.  Thus, we do not eliminate models with any bar-stream interactions.  However, the presence of symmetric and continuous arms precludes models in which true gaps form from interactions with the bar.

	The extended tails can improve constraints on the Galactic potential.  In particular, the tails are sensitive to the shape of the Galactic halo \citep{Bovy:2016cx,Pearson:2015} and longer tails are more constraining.  \cite{Banik:2019} suggest that \palfive{} is a poor candidate for dark matter subhalo searches.  However, the extended tidal tails will afford the opportunity to search for overdensities characteristic of epicycle bunching \cite{Kupper2012} and bar-induced underdensities.  Follow-up observations are needed to confirm the potential tail stars are chemically and kinematically compatible with the cluster.


\section*{Acknowledgments}

JB and JJW received support from the Natural Sciences and Engineering Research Council of Canada (NSERC; funding reference number RGPIN-2015-05235) and from an Ontario Early Researcher Award (ER16-12-061).

This work presents results from the European Space Agency (ESA) space mission Gaia. Gaia data are being processed by the Gaia Data Processing and Analysis Consortium (DPAC). Funding for the DPAC is provided by national institutions, in particular the institutions participating in the Gaia MultiLateral Agreement (MLA). The Gaia mission website is \url{https://www.cosmos.esa.int/gaia}. The Gaia archive website is \url{https://archives.esac.esa.int/gaia}.


\bibliographystyle{mnras}
\bibliography{pal5_in_gaia_dr2_paper}



\appendix

\section{The N-Body Simulation} 
\label{sub:the_nbody}
        
        We use an \nbody{} of \palfive{} to inform the tidal tail search.
        The N-Body simulation uses the the direct $N$-body code \verb!NBODY6!  \citep{Aarseth2006} to simulate the evolution of a \palfive{}-like globular cluster.  The initial cluster is taken to be a Plummer model consisting of 100,000 stars and has a half-mass radius of 10 pc.  Stellar masses follow a Kroupa initial mass function \citep{Kroupa:2001ki}, with the minimum and maximum stellar mass set to $0.1 M_\odot$ and $50 M_\odot$ respectively.  Single stars evolve using the stellar evolution prescription of \citet{hurley2000} assuming a metallicity of Z=0.001 while binary stars, in the event that binaries form, follow \citet{hurley2002}.
    
        The properties of the external tidal field were set to reflect \verb!MWPotential2014!, from \citet{bovy2015}, and is a good approximation of the Galactic potential \citep{Bovy:2016cx}.  The specific linear combination of potentials is a spherical potential from a power-law density with an exponential cut-off Galactic bulge, an NFW dark matter halo \citep{nfw1996}, and a \citet{Miyamoto:1975} disc.  The model cluster was evolved for 12 Gyr with an initial position and velocity that resulted in it being located at the present day location of \palfive{} at the end of the simulation.  The final mass and mass function of the model cluster are comparable to the observed properties of \palfive{} \citep{Ibata2017, grillmair2001}, but the model cluster is too compact relative to \palfive{}, which is in the process of dissolving.
        Changes in the internal dynamics can manifest as differences in the proper motion of tidal tail stars \citep{Ibata2019}; however, small deviations of the potential model to the actual Galactic potential have a more pronounced effect.  As the simulation is primarily used to constrain the kinematics of escaped stars (not the cluster's properties at formation), it was not necessary to reproduce \palfive{} itself exactly.


\section{Constructing the Coordinate System} 
\label{sec:constructing_the_coordinate_system}

	To locally-linearize the \palfive{} tidal tail about the cluster we construct an ICRS sky-rotated coordinate system, with on-sky angles $\phi_1$ and $\phi_2$.  It is constructed as follows: first, select a point along the path of the \nbody{} as the origin; second, using the path of the \nbody{}, rotate and translate the coordinate system until the leading arm is oriented along the positive $\phi_1$ axis.  We found the Cartesian rotation matrix from ICRS to sky-rotated coordinates to be:
	\begin{equation}\label{eq:phi_rotm}
	  \begin{bmatrix}
	    -0.656 & -0.755 & -0.002 \\
	    -0.537 &  0.468 & -0.701 \\
	    0.530 & -0.459 & -0.713
	  \end{bmatrix}
	\end{equation}.

	This rotation places the origin at the cluster and Pole of this coordinate system at ($\alpha_{ngp}, \delta_{ngp}) = (319.27 \degree, -48.31 \degree)$.
	In this sky rotation, the on-tail angle is $\phi_1$ and the orthogonal angle is $\phi_2$.  For the data window, we select a $|\phi_1| < 10 \degree$ by $|\phi_2| < 5 \degree$ data window centered on the Palomar 5 cluster as well points along the \nbody{}.  The data is stitched together to form a continuous window along the \nbody{}'s path.  The selected data window gives a large length of tidal tail, as well as background, to study.


\newpage
\onecolumn
\section{Querying Gaia Database} 
\label{sec:querying_gaia_database}

    Queries to the \href{http://sci.esa.int/gaia/}{\gaia{} database} are done in ADQL.  We provide here the query used to retrieve a dataset containing \palfive\ and a large area of background.  The size of the dataset exceeds the download limit of the \gaia{} database and must be performed on a local version of the data, as described in the  \href{https://github.com/jobovy/gaia_tools}{{\lstinline[
    ]|gaia_tools|} package}.  This query was constructed using the {\lstinline[
    ]|make_query|} function of that package.

\hspace{-3mm}
\begin{minipage}[t]{\dimexpr.5\textwidth-.5\columnsep}
\begin{lstlisting}[language=SQL]
SELECT
gaia.source_id AS id,
gaia.parallax AS prlx, gaia.parallax_error AS prlx_err,
gaia.ra, gaia.ra_error AS ra_err,
gaia.dec, gaia.dec_error AS dec_err,
gaia.pmra, gaia.pmra_error AS pmra_err,
gaia.pmdec, gaia.pmdec_error AS pmdec_err,
ps1.g_mean_psf_mag AS g, ps1.g_mean_psf_mag_error AS g_err,
ps1.r_mean_psf_mag AS r, ps1.r_mean_psf_mag_error AS r_err,
gaia.phi1, gaia.phi2, gaia.pmphi1, gaia.pmphi2

FROM (
    SELECT
    gaia.source_id,
    gaia.parallax, gaia.parallax_error,
    gaia.ra, gaia.ra_error,
    gaia.dec, gaia.dec_error,
    gaia.pmra, gaia.pmra_error,
    gaia.pmdec, gaia.pmdec_error,
    gaia.sinphi1cosphi2, gaia.cosphi1cosphi2, gaia.sinphi2,
    gaia.phi1, gaia.phi2, gaia.c1, gaia.c2,
    ( c1*pmra+c2*pmdec)/cos(phi2) AS pmphi1,
    (-c2*pmra+c1*pmdec)/cos(phi2) AS pmphi2

    FROM (
        SELECT
        gaia.source_id,
        gaia.parallax, gaia.parallax_error,
        gaia.ra, gaia.ra_error,
        gaia.dec, gaia.dec_error,
        gaia.pmra, gaia.pmra_error,
        gaia.pmdec, gaia.pmdec_error,
        gaia.cosphi1cosphi2, gaia.sinphi1cosphi2, gaia.sinphi2,
        gaia.c1, gaia.c2,

        atan2(sinphi1cosphi2, cosphi1cosphi2) AS phi1,
        atan2(sinphi2, sinphi1cosphi2 / sin(atan2(sinphi1cosphi2, cosphi1cosphi2))) AS phi2
        
        FROM (
            SELECT
\end{lstlisting}
\end{minipage}%
\begin{minipage}[t]{\dimexpr.5\textwidth-.5\columnsep}
\begin{lstlisting}[language=SQL]
            gaia.source_id,
            gaia.parallax, gaia.parallax_error,
            gaia.ra, gaia.ra_error,
            gaia.dec, gaia.dec_error,
            gaia.pmra, gaia.pmra_error,
            gaia.pmdec, gaia.pmdec_error,

           -0.6558*cos(radians(dec))*cos(radians(ra))+
            -0.7549*cos(radians(dec))*sin(radians(ra))+
            -0.0022*sin(radians(dec)) AS cosphi1cosphi2,
            
            -0.5373*cos(radians(dec))*cos(radians(ra))+
            0.4688*cos(radians(dec))*sin(radians(ra))+
            -0.7011*sin(radians(dec)) AS sinphi1cosphi2,
            
            0.5303*cos(radians(dec))*cos(radians(ra))+
            -0.4586*cos(radians(dec))*sin(radians(ra))+
            -0.7131*sin(radians(dec)) AS sinphi2,

            -0.9243*cos(radians(dec))-
            0.3817*sin(radians(dec))*cos(radians(ra-
            342.0008)) as c1,
            0.3817*sin(radians(ra-342.0008)) as c2

            FROM gaiadr2.gaia_source AS gaia
        ) AS gaia
    ) AS gaia
) AS gaia

--Comparing to Pan-STARRS1
INNER JOIN gaiadr2.panstarrs1_best_neighbour AS panstarrs1_match ON panstarrs1_match.source_id = gaia.source_id
INNER JOIN gaiadr2.panstarrs1_original_valid AS ps1 ON ps1.obj_id = panstarrs1_match.original_ext_source_id

WHERE
    phi1 > -0.4363
AND phi1 < +0.4363
AND phi2 > -0.1745
AND phi2 < +0.3491
AND prlx < 1

ORDER BY
gaia.source_id
\end{lstlisting}
\end{minipage}



\label{lastpage}\FloatBarrier
\end{document}